\documentclass[reprint,aps,prb,amsmath,amssymb,groupedaddress]{revtex4-2}
\usepackage{graphicx}
\usepackage{dcolumn}
\usepackage{bm}
\usepackage{color} 
\usepackage[tight]{subfigure}
\usepackage{hyperref}

\newcommand{\abs}[1]{\left|{#1}\right|}
\newcommand{\ket}[1]{\left|{}#1 \right>}
\newcommand{\bra}[1]{\left<{}#1 \right|}

\newcommand{\interproduct}[2]{\langle {}#1 | {}#2 \rangle}

\newcommand{\rmi}{\mathrm{i}}
\newcommand{\rme}{\mathrm{e}}

\begin{document}
	
	\title{Periodic jumps in binary lattices with a static force}
	\author{Liwei Duan}
	\email{duanlw@gmail.com}
	\affiliation{Department of Physics, Zhejiang Normal University, Jinhua 321004, China}
	
	\date{\today}
	
	\begin{abstract}
		We investigate the dynamics of a particle in a binary lattice with staggered on-site energies. An additional static force is introduced which further adjusts the on-site energies. The binary lattice appears to be unrelated to the semiclassical Rabi model, which describes a periodically driven two-level system. However, in a certain parity subspace, the Floquet Hamiltonian of the semiclassical Rabi model can be exactly mapped to that of the binary lattice. These connections provide a different perspective for analyzing lattice systems. At resonance, namely that the mismatch of on-site energies between adjacent sites is nearly multiple of the strength of the static force, the level anticrossing occurs. This phenomenon is closely related to the Bloch-Siegert shift in the semiclassical Rabi model. At the $n$th order resonance, an initially localized particle exhibits periodic jumps between site $0$ and site $(2n+1)$, rather than continuous hopping between adjacent sites. The binary lattice with a static force serves as a bridge linking condensed matter physics and quantum optics, due to its connection with the semiclassical Rabi model.
	\end{abstract}
	
	\maketitle
	
	\section{Introduction}
	
	The propagation of a particle in periodic potentials  is a fundamental problem in quantum mechanics and condensed matter physics. Solutions to the Schr\"{o}dinger equation for such systems satisfy Bloch's theorem \cite{Bloch1929}, which yield the periodic Bloch band and delocalized eigenstates. 
	The introduction of an additional static force can profoundly influence the behaviors of the particle, which provides a versatile platform for studying various dynamical behaviors, such as the Bloch oscillation \cite{Hartmann_2004}, 	Bloch-Zener oscillation \cite{Breid_2006} and Rabi oscillation between two Bloch bands \cite{PhysRevB.54.R5235}.
	
	Previous studies on the influence of the static force mainly concentrated on an exact solvable single band approximation, which captures some essential physics in real systems \cite{GLUCK2002103}.
	When a static force is present, the continuous Bloch band transforms into equally spaced discrete energy levels, forming the well-known Wannier-Stark ladder \cite{PhysRev.117.432}. In the meanwhile, the eigenstates become more localized as the strength of the static force increases \cite{GLUCK2002103}.
	The wavepacket exhibits a periodic oscillation, known as the Bloch oscillation, rather than the expected unbounded acceleration towards infinity \cite{Hartmann_2004}. 
	The Bloch oscillations are rarely observable in conventional bulk solids due to the much longer Bloch period compared to the electron scattering time caused by lattice defects \cite{LEO1992943,PhysRevLett.70.3319,PhysRevB.46.7252}. However, it has been experimentally observed in various artificially physical systems, such as the semiconductor superlattice \cite{PhysRevLett.70.3319,LEO1992943,PhysRevB.46.7252}, ultracold atoms in an optical potential \cite{PhysRevLett.76.4508}, waveguide arrays and photonic crystals \cite{PhysRevLett.83.4752,PhysRevLett.83.4756,PhysRevLett.91.263902}, acoustic-cavity structures \cite{PhysRevB.75.024301,PhysRevLett.98.134301} and even a Bose liquid without built-in periodicity \cite{doi:10.1126/science.aah6616}.
	Recently, a form of energy Bloch oscillations is proposed for a periodically driven quantum system characterized by evenly spaced adiabatic energy levels \cite{PhysRevA.98.053820}. In this case, the system's energy will be oscillating, instead of exhibiting a typical real-space oscillation.
	
	Under specific conditions, such as a strong external field, the tunneling between Bloch bands becomes non-negligible \cite{doi:10.1098/rspa.1934.0116,GLUCK2002103}, exceeding the capabilities of the single-band approximation. A binary lattice, described by the period-doubled tight-binding model, possesses two Bloch bands and serves as one of the simplest platforms to investigate the interband tunneling effect \cite{Xian-Geng_Zhao_1992}. The competitions between the Bloch oscillation and the interband tunneling lead to the Bloch-Zener oscillation
	\cite{Breid_2006,Breid_2007,Martin_Holthaus_2000,PhysRevLett.74.1831}, which has also been observed in the waveguide-based superlattice \cite{PhysRevLett.102.076802}.
	The Bloch-Zener oscillation paves a way to perform quantum walks \cite{PhysRevA.82.033602,Longhi_2012} and generate widely tunable matter wave beam splitters and Mach–Zehnder interferometers \cite{Breid_2007}.
	
	Recently, in quantum optical systems, the concept of a Fock-state lattice has emerged, where the latticelike structure emerges by identifying the different Fock states as the lattice sites \cite{PhysRevA.108.033721,doi:10.1126/science.ade6219,10.1093/nsr/nwaa196}. As a paradigmatic model in quantum optics, the quantum Rabi model describes the simplest interaction between a two-level atom and a quantized light field. It has been mapped into a Fock-state lattice to explore a different type of topological phases arising from quantized light \cite{doi:10.1126/science.ade6219,10.1093/nsr/nwaa196} and amplitude-modulated Bloch oscillations \cite{Zhang_2015}. The semiclassical Rabi model, on the other hand, describes a two-level atom driven by a periodic classical light field \cite{PhysRev.51.652,Braak_2016}. It cannot be mapped into a Fock-state lattice due to the classical field. Nevertheless, the time-dependent semiclassical Rabi model can be transformed into a time-independent one with an infinite-dimensional Hilbert space according to Floquet's theory \cite{PhysRev.138.B979,doi:https://doi.org/10.1002/9783527624003.ch2,GRIFONI1998229}. The Floquet states can be regarded as the lattice sites, which provide an opportunity to create a latticelike structure. The latticelike structure formed by the Floquet states may, in a sense, be reminiscent of the Floquet topological systems \cite{PhysRevB.82.235114,https://doi.org/10.1002/pssr.201206451,Rechtsman2013}, which enhance the flexibility of the Hamiltonian and broaden the general classification of topological phases by introducing the periodicity in the time domain. Nonetheless, significant distinctions exist. We illustrate that a basic two-level system exhibits a latticelike structure under periodic driving, whereas the systems they studied constitute a lattice even in the absence of driving.
	
	In this paper, we investigate the correspondence between the binary lattice subjected to a static external force and the semiclassical Rabi model. Our primary focus is on the periodic jumps within the binary lattice, a phenomenon closely associated with resonance phenomena and the Bloch-Siegert shift in the semiclassical Rabi model.	
	The paper is structured as follows. In Sec. \ref{sec:binary}, we introduce the Hamiltonian of the binary lattice with a static force. In Sec. \ref{sec:Rabi}, we provide a brief overview of the Floquet Hamiltonian of the semiclassical Rabi model and introduce a parity operator that divides the entire Hilbert space into two distinct subspaces with even and odd parities, respectively. We then demonstrate the exact equivalence between the Floquet Hamiltonian of the semiclassical Rabi model and the Hamiltonian of the binary lattice. The development of various approaches and the discovery of numerous phenomena in the semiclassical Rabi model can be readily extended to that in the binary lattice. In Sec. \ref{sec:results}, we present the level anticrossing at the resonance, as well as the periodic jumps between different sites. Finally, a brief summary is given is Sec. \ref{sec:conclusion}.
	
	\section{Binary lattice with a static force} \label{sec:binary}
	
\begin{figure}[htb]
		\centering
		\includegraphics[scale=1]{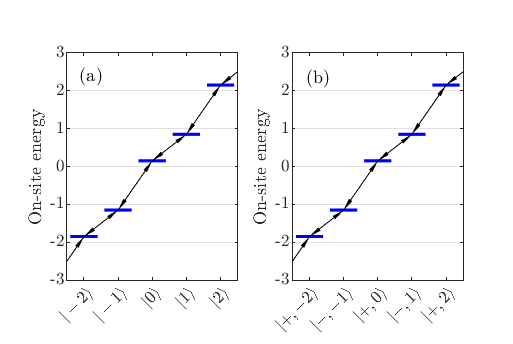} 
		\caption{Level structures for (a) the binary lattice and (b) the odd parity chain of the semiclassical Rabi model. Each lattice site is represented by the basis state along the horizontal axis, with the blue line indicating the respective on-site energy. The black arrows denote the hopping between adjacent sites. The other parameters are $F=\omega=1$ and $\epsilon= \Omega = 0.3$.}\label{fig:WSL}
	\end{figure}

	In this paper, we consider a tight-binding model that describes a binary lattice subjected to a static force as follows:
	\begin{eqnarray}\label{eq:H} 
		\hat{H} &=& -V \sum_{n = -\infty}^{+\infty} \left(\ket{n} \bra{n + 1} + \ket{n + 1} \bra{n}\right) \\
		&&+ \sum_{n = -\infty}^{+\infty} \left(F n + \frac{\epsilon}{2} (-1)^n\right) \ket{n} \bra{n} ,\nonumber 
	\end{eqnarray}
	where $\ket{n}$ is the Wannier state localized at site $n$. $V$ and $\epsilon$ denote the hopping rate and on-site energy mismatch between nearest-neighbor sites, respectively, which together give rise to two Bloch bands \cite{Breid_2006}. $F$ corresponds to the external static force. Alternately, Hamiltonian (\ref{eq:H}) can be written in a matrix form as follows:
	\begin{eqnarray} \label{eq:Hm}
		\hat{H} = \left(
		\begin{array}{ccccccc}
			\ddots&\ddots&\ddots&&&&\\
			&-V& -\frac{\epsilon}{2} - F & -V&&&\\
			&&-V& \frac{\epsilon}{2} & -V&&\\
			&&&-V& -\frac{\epsilon}{2} + F & -V&\\
			&&&&\ddots&\ddots&\ddots\\
		\end{array}
		\right).
	\end{eqnarray}
	The corresponding level structure is shown in Fig. \ref{fig:WSL}(a).
	In the absence of the energy mismatch $\epsilon$, Eq. (\ref{eq:H}) reduces to the famous Wannier-Stark Hamiltonian, whose eigenenergies take the form of the Wannier-Stark ladder \cite{PhysRev.117.432,Hartmann_2004}.
	
	For clarity, we can introduce three operators $\hat{E}_0$ and $\hat{E}_{\pm}$, as follows \cite{KORSCH200354,Hartmann_2004,Breid_2006}:
	\begin{subequations}\label{eq:EO}
		\begin{eqnarray} 
			\hat{E}_0 &=& \sum_{n = -\infty}^{+\infty} n \ket{n} \bra{n}, \\
			\hat{E}_+ &=& \sum_{n = -\infty}^{+\infty} \ket{n + 1} \bra{n},\\
			\hat{E}_- &=& \sum_{n = -\infty}^{+\infty} \ket{n} \bra{n + 1},
		\end{eqnarray}
	\end{subequations}
	which correspond to the generators of Euclidean algebra, satisfying the following commutation relations \cite{doi:https://doi.org/10.1002/9783527624003.ch2}:
	\begin{subequations}
	\begin{eqnarray}
		\left[\hat{E}_0, \hat{E}_{\pm}\right] &=& \pm \hat{E}_{\pm}, \\
		\left[\hat{E}_+, \hat{E}_-\right] &=& 0 .
	\end{eqnarray}
	\end{subequations}
	It is important to note that the Wannier state $\ket{n}$ is the eigenstate of $\hat{E}_0$ with eigenvalue $n$. Additionally, $\hat{E}_{\pm}$ act as raising and lowering operators, respectively, as indicated by
	\begin{eqnarray}
		\hat{E}_{\pm} \ket{n} = \ket{n \pm 1} .
	\end{eqnarray}
	In terms of $\hat{E}_0$ and $\hat{E}_{\pm}$, Hamiltonian (\ref{eq:H}) can be rewritten as
	\begin{eqnarray} \label{eq:He}
		\hat{H} = - V \left(\hat{E}_+ + \hat{E}_-\right) + F \hat{E}_0 + \frac{\epsilon}{2} (-1)^{\hat{E}_0} .
	\end{eqnarray}

	\section{Relations between the binary lattice and the semiclassical Rabi model}\label{sec:Rabi}
	
	The semiclassical Rabi model, serving as a prototype in quantum optics, has consistently attracted attention since its inception \cite{PhysRev.51.652,Braak_2016,GRIFONI1998229}. Its Hamiltonian can be expressed as
	\begin{eqnarray}
		\hat{H}(t) = \frac{\Omega}{2} \hat{\sigma}_z - 2 \lambda \hat{\sigma}_x \cos \omega t ,
	\end{eqnarray}
	where $\hat{\sigma}_{x,y,z}$ represent the Pauli matrices which are employed to describe the two-level system. $\Omega$ denotes the energy difference of the two-level system, $\omega$ is the frequency of the classical light field, and $\lambda$ stands for the coupling strength between them.
	
	According to Floquet's theory \cite{PhysRev.138.B979,doi:https://doi.org/10.1002/9783527624003.ch2}, the time-dependent Hamiltonian can be replaced by a time-independent counterpart with an infinite-dimensional Hilbert space as follows:
	\begin{eqnarray}
		\hat{\mathcal{H}}_F = \frac{\Omega}{2} \hat{\sigma}_z + \omega \hat{E}_0 - \lambda \hat{\sigma}_x \left(\hat{E}_+ + \hat{E}_-\right) . \label{eq:HF}
	\end{eqnarray}
	$\hat{E}_0$ and $\hat{E}_{\pm}$ are given by Eq. (\ref{eq:EO}), with $n$ now corresponding to the Fourier exponent.
	
	The Floquet Hamiltonian (\ref{eq:HF}) is of infinite dimensions, whose exact analytical solutions have remained elusive up to now. Nevertheless, its dimensions can be reduced by exploiting its symmetry. We begin by introducing a parity operator defined as
	\begin{eqnarray}
		\hat{\Pi} &=& \exp\left[\rmi \pi \left(\hat{\sigma}_+ \hat{\sigma}_- + \hat{E}_0\right)\right] \\
		&=& -\hat{\sigma}_z (-1)^{\hat{E}_0} ,\nonumber
	\end{eqnarray}
	with $\hat{\sigma}_{\pm} = \left(\hat{\sigma}_x \pm \rmi \hat{\sigma}_y\right) / 2$. It can be easily demonstrated that $\hat{\Pi} \hat{\mathcal{H}}_F \hat{\Pi}^{\dagger} = \hat{\mathcal{H}}_F$, which indicates that the Floquet Hamiltonian $\hat{\mathcal{H}}_F$ admits the parity symmetry. The parity operator $\hat{\Pi}$ possesses eigenvalues $\Pi = \pm 1$, which separate the whole Hilbert space into two independent subspaces characterized by even and odd parities respectively. These are commonly referred to as the parity chains \cite{PhysRevLett.105.263603,PhysRevLett.108.163601}, illustrated as follows:
	\begin{eqnarray}
		&\dots \leftrightarrow \ket{+,-1} \leftrightarrow \ket{-,0} \leftrightarrow \ket{+,1} \leftrightarrow \dots \left(\Pi = +1\right), \\
		&\dots \leftrightarrow \ket{-,-1} \leftrightarrow \ket{+,0} \leftrightarrow \ket{-,1} \leftrightarrow \dots \left(\Pi = -1\right),
	\end{eqnarray}
	where the basis state is $\ket{s,n}=\ket{s}\ket{n}$ with $\hat{\sigma}_z \ket{s} = s \ket{s}$ ($s=\pm$) and $\ket{n}$ the Floquet states.
	
	In the basis of $\left\{\ket{s,n}\right\}$, the matrix elements of the Floquet Hamiltonian are given by
	\begin{eqnarray}
		\bra{s,n} \hat{\mathcal{H}}_F \ket{s',n'} &=& \left(s \frac{\Omega}{2} + \omega n\right) \delta_{s,s'} \delta_{n,n'}\\
		&& - \lambda \delta_{s,-s'} \delta_{n, n' \pm 1}.\nonumber
	\end{eqnarray}
	In the odd parity subspace ($\Pi = -1$), the matrix form of the Floquet Hamiltonian can be written as
	\begin{eqnarray}\label{eq:H-}
		\hat{\mathcal{H}}_{-} = \left(
		\begin{array}{ccccccc}
			\ddots&\ddots&\ddots&&&&\\
			&-\lambda& -\frac{\Omega}{2} - \omega & -\lambda&&&\\
			&&-\lambda& \frac{\Omega}{2} & -\lambda&&\\
			&&&-\lambda& -\frac{\Omega}{2} + \omega & -\lambda&\\
			&&&&\ddots&\ddots&\ddots\\
		\end{array}
		\right).
	\end{eqnarray}
	A transformation of $\Omega$ to $-\Omega$ results in the Floquet Hamiltonian matrix for the even parity subspace ($\Pi = 1$). Obviously, $\hat{\mathcal{H}}_-$ [Eq. (\ref{eq:H-})] is exactly the same as $\hat{H}$ [Eq. (\ref{eq:Hm})], as long as we choose $\omega = F$, $\Omega = \epsilon$ and $\lambda = V$.
	Inspired by the Fock-state lattice \cite{PhysRevA.108.033721}, one can interpret the diagonal elements of the Floquet Hamiltonian as on-site energies of the lattice. Meanwhile, the off-diagonal elements represent the hopping rates between these sites, as illustrated in Fig. \ref{fig:WSL} (b). 
	An alternate approach is presented in Appendix \ref{sec:equivalence}, which utilizes the Fulton-Gouterman transformation to establish the equivalence of the Hamiltonians between the binary lattice and the semiclassical Rabi model. It is important to note that the time evolution of two models does not exhibit a simple and straightforward correspondence as that of the Hamiltonians, as discussed in Appendix \ref{sec:difference}. Nevertheless, the quasienergy of the semiclassical Rabi model and the eigenenergy of the binary lattice are comparable; so are the corresponding eigenstates. From them, some dynamical behaviors are predictable, such as the periodic jump.
				
	\section{Results and discussions} \label{sec:results}
	
	As demonstrated in Sec. \ref{sec:Rabi}, the Hamiltonian matrix of the binary lattice with a static force is equivalent to that of the semiclassical Rabi model in the odd parity subspace. Consequently, analytical and numerical solutions developed for the semiclassical Rabi model can be readily extended to those for the binary lattice, and vice versa.
	
	One of the most studied phenomena in the semiclassical Rabi model is the Bloch-Siegert shift \cite{PhysRev.57.522,PhysRev.138.B979,PhysRevLett.105.257003,doi:https://doi.org/10.1002/9783527624003.ch2,PhysRevA.86.023831}. The resonance between the two-level system and the driven light field does not occur exactly at $\Omega = (2n + 1) \omega$ with $n=0,1,2,\dots$ The shift in resonance frequency, denoted as $\delta = (2n + 1)\omega - \Omega$, is termed the Bloch-Siegert shift and can be determined by the position of the level anticrossing point \cite{PhysRev.138.B979,doi:https://doi.org/10.1002/9783527624003.ch2}. Similar phenomena are expected to occur in the case of the binary lattice subjected to a static field near the $n$th order resonance, specifically when $\epsilon \approx (2n + 1) F$.
	
	For simplicity, we begin by examining the zeroth order resonance of the binary lattice. As indicated by Eq. (\ref{eq:Hm}), the Wannier states $\ket{0}$ and $\ket{1}$ are degenerate for $\epsilon = F$. Nonetheless, the hopping between them, represented by the off-diagonal element of the Hamiltonian matrix, denoted as $V$, breaks this degeneracy. According to the degenerate perturbation theory, we obtain a $2\times 2$ effective Hamiltonian matrix as follows:
	\begin{eqnarray} \label{eq:H2}
		\hat{H}_2 = \left(
		\begin{array}{cc}
			\frac{\epsilon}{2} & -V\\
			-V & -\frac{\epsilon}{2} + F
		\end{array}
		\right),
	\end{eqnarray}
	whose eigenenergies and eigenstates are
	\begin{subequations}
	\begin{eqnarray}
		e_{\pm} &=& \frac{F \pm \Delta}{2} ,\label{eq:e2}\\
		\ket{\phi_+} &=& \cos \frac{\theta}{2} \ket{1} - \sin \frac{\theta}{2} \ket{0} ,\\
		\ket{\phi_-} &=& \cos \frac{\theta}{2} \ket{0} + \sin \frac{\theta}{2} \ket{1} ,
	\end{eqnarray}
	\end{subequations}
	with $\Delta=\sqrt{\left(\epsilon - F\right)^2 + 4 V^2}$ and $\theta = \arcsin \frac{2 V}{\Delta}$. 
	Clearly, the gap between two eigenstates is given by $\Delta$. It reaches the minimum at the resonance $\epsilon = F$, which also determines the level anticrossing point. Near resonance, the eigenstates tend to be equally distributed between $\ket{0}$ and $\ket{1}$, while away from resonance they tend to localize on either $\ket{0}$ or $\ket{1}$.
	Near resonance, if the particle is initially localized at $\ket{0}$, it will oscillate between $\ket{0}$ and $\ket{1}$ with a period of $2 \pi / \Delta$. Finally, the probability that the particle transfers to $\ket{1}$ is given by 
	\begin{eqnarray}
		P_{0\rightarrow1} = \frac{4 V^2}{\Delta^2} \sin^2 \left(\frac{\Delta t}{2}\right) .
	\end{eqnarray}
	The amplitude of the oscillation is largest at resonance when $F=\epsilon$.
	
	\begin{figure}[htb]
		\centering
		\includegraphics[scale=1]{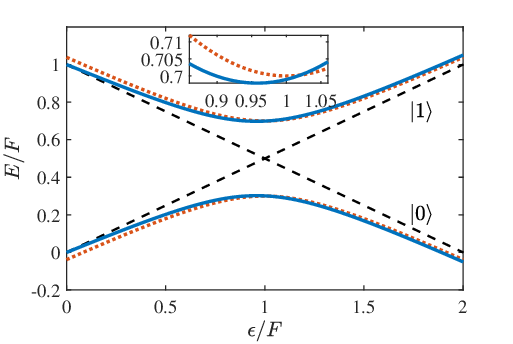} 
		\caption{Eigenenergies as a function of $\epsilon/F$ at $V/F=0.2$. Blue solid lines represent the numerical exact results, while the red dotted lines correspond to the perturbative analytical results from Eq. (\ref{eq:e2}). The black dashed lines represent the on-site energies of $\ket{0}$ and $\ket{1}$.  The inset provides a detailed view of the upper branch near the level anticrossing. }\label{fig:BSS}
	\end{figure}

	Figure \ref{fig:BSS} displays the eigenenergies of $\hat{H}_2$ [Eq. (\ref{eq:H2})], as well as the corresponding numerical exact results of $\hat{H}$ [Eq. (\ref{eq:Hm})]. Although we can confirm the existence of the level anticrossing at $\epsilon/F=1$ from $\hat{H}_2$, it fails to predict the Bloch-Siegert shift, as depicted in the inset. For $V/F=0.2$, numerical exact results indicate that the resonance or level anticrossing occurs at $\epsilon/F \approx 0.9579$, with a corresponding energy gap of $\Delta_{\text{min}}/F \approx 0.3958$.
	
	\begin{figure}[htb]
		\centering
		\includegraphics[scale=1]{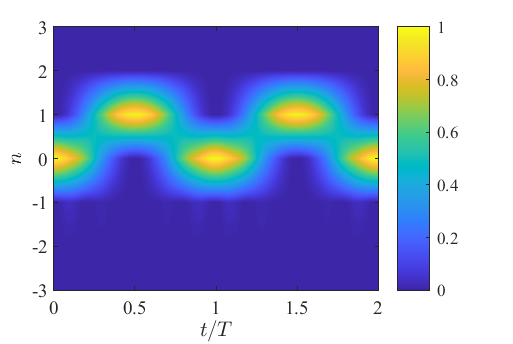} 
		\caption{Dynamics of the probability distribution $P_n(t)$ at the zeroth order resonance with $\epsilon/F=0.9579$ and $V/F=0.2$. The particle is initially located at site $0$. }\label{fig:dynamics1}
	\end{figure}
	
	Here we concentrate on the dynamics at resonance. Without loss of generality, we assume the initial state to be $\ket{\psi(0)} = \ket{0}$. The probability of finding the particle at site $n$ is given by $P_n(t) = \abs{\interproduct{n}{\psi(t)}}^2$, as shown in Fig. \ref{fig:dynamics1}. The dynamical behavior aligns with our earlier perturbative analysis, specifically, it exhibits periodic oscillations between sites $0$ and $1$ with a period of $T=2\pi/\Delta_{\min}$.
	
	\begin{figure}[htb]
		\centering
		\includegraphics[scale=1]{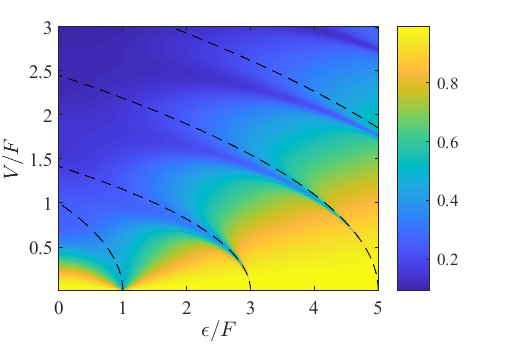} 
		\caption{IPR as a function of $V$ and $\epsilon$. The dashed lines correspond to the resonant condition determined using the Bloch-Siegert shift derived by Shirley \cite{PhysRev.138.B979}.}\label{fig:IPR}
	\end{figure}
	
	To study the higher order resonance, we introduce the inverse participation ratio (IPR) \cite{Wegner1980}, which is defined as
	\begin{eqnarray}
		\text{IPR} = \frac{\sum_{n=-\infty}^{+\infty} \abs{\interproduct{n}{\phi}}^4}{\left(\sum_{n=-\infty}^{+\infty} \abs{\interproduct{n}{\phi}}^2\right)^2},
	\end{eqnarray}
	where $\ket{\phi}$ represents an arbitrary eigenstate of $\hat{H}$ [Eq. (\ref{eq:Hm})]. As demonstrated in Ref. \cite{Breid_2006}, different eigenstates can be transformed into each other by translation and inversion operators, which do not influence the IPR. The IPR as a function of $V$ and $\epsilon$ is shown in Fig. \ref{fig:IPR}. In general, IPR tends to decrease with an increase in the hopping rate $V$, suggesting that the eigenstates tend to become delocalized. On the contrary, IPR tends to increase with an increase in the on-site energy mismatch $\epsilon$, indicating that the eigenstates tend to become localized. Therefore, when $V$ is small and $\epsilon$ is large, the eigenstate tends to become localized with IPR$ \rightarrow 1$ as shown by the yellow region in the lower-right corner of Fig. \ref{fig:IPR}. However, particular attention should be paid  to the vicinity of the resonance $\epsilon \approx (2n + 1) F$. The eigenstates tend to be the superposition of two nearly degenerate states, which leads to IPR$ \rightarrow 1/2$ at resonance. Given that the energy mismatch required to attain the $n$-th order resonance is $\epsilon = (2n + 1)F - \delta$, $\delta$ corresponds to the Bloch-Siegert shift in the semiclassical Rabi model.
	Shirley determined the Bloch-Siegert shift in the semiclassical Rabi model by Salwen's perturbation theory \cite{PhysRev.138.B979}, which can also be employed to describe the current model,
	\begin{eqnarray}
		\delta = \left\{
			\begin{array}{cc}
				\frac{V^2}{F}, & \text{ for } n=0 ,\\
				\frac{2n + 1}{n(n + 1)} \frac{V^2}{F}, & \text{ for } n \ge 1 .
			\end{array}
			\right.
	\end{eqnarray}
	The dashed line in Fig. \ref{fig:IPR} corresponds to the resonant condition by employing the Bloch-Siegert shift derived by Shirley, which is consistent with the numerical results, especially for $V/F \ll 1$.
	
	\begin{figure}[htb]
		\centering
		\includegraphics[scale=1]{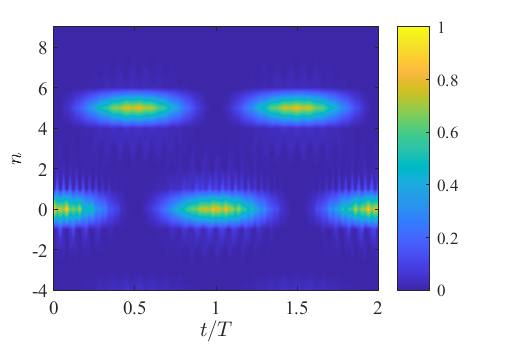} 
		\caption{Dynamics of the probability distribution $P_n(t)$ at the second order resonance with $\epsilon/F=4.11467$ and $V/F=1$. The particle is initially located at site $0$. }\label{fig:dynamics2}
	\end{figure}

	Numerical calculation indicates that the second order resonance occurs at $\epsilon/F \approx 4.11467$ for $V/F=1$ and the corresponding energy gap is $\Delta_{\min}/F \approx 0.03208$. The dynamics of the probability distribution $P_n(t)$ is shown in Fig. \ref{fig:dynamics2}. Instead of continuous transfer between adjacent sites, the dynamics shows a periodic jump between site $0$ and site $5$ with a period of $T=2\pi/\Delta_{\min}$. At the $n$th order resonance, we expect that the periodic jump between site $0$  and site $(2n + 1)$ will occur.
	
	\section{Conclusions} \label{sec:conclusion}
	
	In this paper, we conducted both analytical and numerical investigations of a binary lattice subjected to a static external force. We began by establishing the connections between the binary lattice and the semiclassical Rabi model -- a periodic driving two-level system: the Floquet Hamiltonian of the semiclassical Rabi model within a specific parity subspace is precisely equivalent to the Hamiltonian of the binary lattice subjected to a static force. Consequently, solutions derived for the semiclassical Rabi model can be readily extended to the binary lattice and vice versa.
	
	Here we concentrated  on the resonance and level anticrossing phenomena in the binary lattice subjected to a static force, which are closely related with the Bloch-Siegert shift observed in the semiclassical Rabi model. At the $n$th order resonance [$\epsilon \approx (2n + 1) F$], the eigenstates tend to be a superposition of Wannier states $\ket{0}$ and $\ket{2n+1}$, while becoming localized on one of the Wannier states away from the resonance. This phenomenon can be confirmed through the IPR. When a particle initially resides at site $0$, it presents a periodic jump between site $0$ and site $(2n+1)$, rather than a continuous hopping between adjacent sites. The period of jumps is determined by the energy gap. 
	
	The correspondence between the binary lattice subjected to a static force and the semiclassical Rabi model provides insights into bridging condensed matter physics and quantum optics.

	\appendix
	
	\section{Equivalence of the Hamiltonians between the binary lattice and the semiclassical Rabi model} \label{sec:equivalence}
	
	In the basis state of $\ket{\pm x} = \left(\ket{+} \pm \ket{-}\right) / \sqrt{2}$ which satisfy $\hat{\sigma}_x \ket{\pm x} = \pm \ket{\pm x}$, the Floquet Hamiltonian (\ref{eq:HF}) of the semiclassical Rabi model can be rewritten in a matrix form as follows:
	\begin{equation} \label{eq:H_matrix}
		\hat{\mathcal{H}}_F = \left(
		\begin{array}{cc}
			\omega \hat{E}_0 - \lambda \left(\hat{E}_+ + \hat{E}_-\right) & -\frac{\Omega}{2} \\
			-\frac{\Omega}{2} & \omega \hat{E}_0 + \lambda \left(\hat{E}_+ + \hat{E}_-\right)
		\end{array}
		\right) .
	\end{equation}
	Furthermore, we can introduce the Fulton-Gouterman transformation \cite{doi:10.1063/1.1701181,PhysRevLett.107.100401},
	\begin{eqnarray}
		\hat{U} = \frac{1}{\sqrt{2}} \left(
		\begin{array}{cc}
			1 & 1\\
			(-1)^{\hat{E}_0} & -(-1)^{\hat{E}_0}
		\end{array}
		\right) ,
	\end{eqnarray}
	with which Eq. (\ref{eq:H_matrix}) can be transformed into a diagonal form, namely, $\hat{U}^{\dagger} \hat{\mathcal{H}}_F \hat{U} = \text{diag} \left(\hat{H}_+, \hat{H}_-\right)$. $\hat{H}_+$ and $\hat{H}_-$ correspond to even ($\Pi = 1$) and odd ($\Pi = -1$) parities, respectively. They are defined as
	\begin{eqnarray} \label{eq:H_parity}
		\hat{H}_{\pm} &=& \omega \hat{E}_0 - \lambda \left(\hat{E}_+ + \hat{E}_-\right) \mp \frac{\Omega}{2} (-1)^{\hat{E}_0} \\
		&=& \sum_{n = -\infty}^{+\infty} \left(\omega n \mp \frac{\Omega}{2} (-1)^n\right) \ket{n} \bra{n} \nonumber\\
		&& - \lambda \left(\ket{n} \bra{n + 1} + \ket{n + 1} \bra{n}\right). \nonumber
	\end{eqnarray}
	It is obvious that the Hamiltonian (\ref{eq:H_parity}) in the odd parity subspace is equivalent to Eqs. (\ref{eq:H}) and (\ref{eq:He}), which is just the Hamiltonian of the binary lattice subjected to a static force.
	
	\section{Differences in the time evolution between the binary lattice and the semiclassical Rabi model}\label{sec:difference}
	
	For the binary lattice subjected to a static force described by Hamiltonian (\ref{eq:H}), we can assume that one of the eigenstates is denoted as
	\begin{eqnarray}
		\ket{\phi_0^{(L)}} = \sum_{n=-\infty}^{+\infty} c_n \ket{n},
	\end{eqnarray}
	with the corresponding eigenenergy $e_0^{(L)}$. It is straightforward to confirm that 
	\begin{eqnarray}
		\ket{\phi_m^{(L)}} = \hat{E}_+^{2 m} \ket{\phi_0^{(L)}} = \sum_{n=-\infty}^{+\infty} c_n \ket{n + 2 m}
	\end{eqnarray}
	are also eigenstates with eigenenergies $e_m^{(L)} = e_0^{(L)} + 2 m F$ ($m=0,\pm 1, \pm2, \dots$) \cite{Breid_2006}, which form an equally spaced energy ladder. Given that the initial state is $\ket{\phi_m^{(L)}}$, the time evolution is governed by the time-dependent wave function
	\begin{eqnarray}
		\ket{\psi_m^{(L)}(t)} = \rme^{-\rmi e_m^{(L)} t} \ket{\phi_m^{(L)}},
	\end{eqnarray}
	which is obviously dependent on $m$.
	
	In the semiclassical Rabi model with Floquet Hamiltonian (\ref{eq:HF}), we can assume that one of the eigenstates is denoted as
	\begin{eqnarray}
		\ket{\phi_0^{(R)}} = \sum_{s=\pm} \sum_{n=-\infty}^{+\infty} c_{s,n} \ket{s,n},
	\end{eqnarray}
	with the corresponding quasienergy $e_0^{(R)}$. Similar to that in the binary lattice, we can also obtain a set of eigenstates written as
	\begin{eqnarray}
		\ket{\phi_m^{(R)}} = \hat{E}_+^{2 m} \ket{\phi_0^{(R)}} = \sum_{s=\pm} \sum_{n=-\infty}^{+\infty} c_{s,n} \ket{s,n + 2 m},
	\end{eqnarray}
	with quasienergies $e_m^{(R)} = e_0^{(R)} + 2 m \omega$. 
	According to Floquet's theory, we need to introduce
	\begin{widetext}
	\begin{eqnarray}
		\ket{\phi_m^{(R)}} = \sum_{s=\pm} \sum_{n=-\infty}^{+\infty} c_{s,n} \ket{s,n + 2 m} \rightarrow \ket{\phi_m^{(R)} (t)} = \sum_{s=\pm} \sum_{n=-\infty}^{+\infty} c_{s,n} \rme^{\rmi (n + 2m) \omega t} \ket{s} .
	\end{eqnarray}
	\end{widetext}
	The time evolution corresponding to $\ket{\phi_m^{(R)}}$ is given by 
	\begin{eqnarray}
		\ket{\psi_m^{(R)}(t)} &=& \rme^{-\rmi e_m^{(R)} t} \ket{\phi_m^{(R)}(t)} \\
		&=& \rme^{-\rmi \left(e_0^{(R)} + 2 m \omega\right) t} \sum_{s=\pm} \sum_{n=-\infty}^{+\infty} \rme^{\rmi \left(n + 2 m\right) \omega t} c_{s,n} \ket{s} \nonumber\\
		&=& \rme^{-\rmi e_0^{(R)} t} \sum_{s=\pm} \sum_{n=-\infty}^{+\infty} \rme^{\rmi n \omega t} c_{s,n} \ket{s} \nonumber\\
		&=& \ket{\psi_0^{(R)}(t)} ,\nonumber
	\end{eqnarray}
	which does not depend on $m$. 
	
	The difference in the time evolution is easy to understand. Despite the infinite-dimensional nature of the Floquet Hamiltonian in the semiclassical Rabi model, the original Hamiltonian is fundamentally two-dimensional, in stark contrast to that of the binary lattice.
	For the basis state $\ket{s,n} = \ket{s} \ket{n}$, the spin component $\ket{s}$ represents the physical state, while the Floquet state $\ket{n}$ serves a purely auxiliary role. The inclusion of the Floquet state $\ket{n}$ is essential for constructing the Floquet Hamiltonian, determining the corresponding quasienergies, and identifying the eigenstates. Nevertheless, it is invisible in the time evolution. Therefore, there exist differences in the time evolution between the binary lattice and the semiclassical Rabi model. Dynamical phenomena observed in the binary lattice with a static force, like Bloch-Zener oscillations, are challenging to detect in the semiclassical Rabi model and vice versa.

	\begin{acknowledgments}
		This research was supported by the National Natural Science Foundation of China (NSFC) under Grant No. 12305032 and Zhejiang Provincial Natural Science Foundation of China under Grant No. LQ23A050003.
	\end{acknowledgments}

\end{document}